\documentclass[sigconf]{acmart}
\AtBeginDocument{%
  \providecommand\BibTeX{{%
    \normalfont B\kern-0.5em{\scshape i\kern-0.25em b}\kern-0.8em\TeX}}}

\usepackage{tabularx}
\usepackage{booktabs}
\usepackage{graphicx}
\usepackage{hyperref}
\usepackage{hhline}

\usepackage{subcaption}
\usepackage{enumitem}
\usepackage{float} 
\usepackage{soul,color}
\usepackage[utf8]{inputenc}

\usepackage{longtable}
\usepackage{chngcntr}
\usepackage{xcolor}
\usepackage{pbox}
\usepackage{multirow}
\usepackage{array}
\usepackage{makecell}
\usepackage{afterpage} 
\usepackage[title]{appendix}
\usepackage{subcaption,graphicx}

\newcolumntype{L}[1]{>{\raggedright\let\newline\\\arraybackslash\hspace{0pt}}m{#1}}
\newcolumntype{C}[1]{>{\centering\let\newline\\\arraybackslash\hspace{0pt}}m{#1}}
\newcolumntype{R}[1]{>{\raggedleft\let\newline\\\arraybackslash\hspace{0pt}}m{#1}}
\copyrightyear{2023}
\acmYear{2023}
\setcopyright{rightsretained}
\acmConference[CHI EA '23]{Extended Abstracts of the 2023 CHI Conference on Human Factors in Computing Systems}{April 23--28, 2023}{Hamburg, Germany}
\acmBooktitle{Extended Abstracts of the 2023 CHI Conference on Human Factors in Computing Systems (CHI EA '23), April 23--28, 2023, Hamburg, Germany}
\acmDOI{10.1145/3544549.3585904}
\acmISBN{978-1-4503-9422-2/23/04}
\begin{document}

%%
%% The "title" command has an optional parameter,
%% allowing the author to define a "short title" to be used in page headers.

% Share link: https://www.overleaf.com/4755158437hpxywbppmdxc
\title[Joint Family Oversight for Mobile Privacy and Security]{It Takes a Village: A Case for Including Extended Family Members in the Joint Oversight of Family-based Privacy and Security for Mobile Smartphones}

%Joint Family Oversight for Mobile Privacy, Safety and Security beyond the Parent-Teen Relationship}
% Co-managing Mobile Online Safety and Privacy with Extended Families
%%% Share link: https://www.overleaf.com/4755158437hpxywbppmdxc

%% The "author" command and its associated commands are used to define
%% the authors and their affiliations.
%% Of note is the shared affiliation of the first two authors, and the
%% "authornote" and "authornotemark" commands
%% used to denote shared contribution to the research.

% \author{ANONYMOUS AUTHOR(S)}
\author{Mamtaj Akter}
\email{Mamtaj.Akter@vanderbilt.edu}
\orcid{0000-0002-5692-9252}
\affiliation{%
  \institution{Vanderbilt University}
  \city{Nashville}
  \state{Tennessee}
  \postcode{37212}
  \country{USA}
}

\author{Leena Alghamdi}
\email{Leenaalghamdi@knights.ucf.edu}
\orcid{0000-0003-2102-9155}
\affiliation{%
  \institution{University of Central Florida}
  \city{Orlando}
  \state{Florida}
  \postcode{32826}
  \country{USA}
}

\author{Jess Kropczynski}
\email{Jess.Kropczynski@uc.edu}
\orcid{0000-0002-7458-6003}
\affiliation{%
  \institution{University of Cincinnati}
  % \streetaddress{P.O. Box 1234}
  \city{Cincinnati}
  \state{OH}
  \postcode{45221}
  \country{USA}
}

\author{Heather Lipford}
\email{Heather.Lipford@uncc.edu}
\orcid{0000-0002-5261-0148}
%\email{author2@email.com}
\affiliation{%
  \institution{University of North Carolina, Charlotte}
  %\streetaddress{P.O. Box 1234}
  \city{Charlotte}
  \state{North Carolina}
  \postcode{28223}
  \country{USA}
 }

\author{Pamela J. Wisniewski}
\email{Pamela.Wisniewski@vanderbilt.edu}
\orcid{0000-0002-6223-1029}
\affiliation{%
  \institution{Vanderbilt University}
  \city{Nashville}
  \state{Tennessee}
  \postcode{37212}
  \country{USA}
}
 
%\authornote{Both authors contributed equally to this research.}

%%
%% By default, the full list of authors will be used in the page
%% headers. Often, this list is too long, and will overlap
%% other information printed in the page headers. This command allows
%% the author to define a more concise list
%% of authors' names for this purpose.
\renewcommand{\shortauthors}{Akter et al.}

%%
%% The abstract is a short summary of the work to be presented in the
%% article.
\begin{abstract}
We conducted a user study with 19 parent-teen dyads to understand the perceived benefits and drawbacks of using a mobile app that allows them to co-manage mobile privacy, safety, and security within their families. While the primary goal of the study was to understand the use case as it pertained to parents and teens, an emerging finding from our study was that participants found value in extending app use to other family members (siblings, cousins, and grandparents). Participants felt that it would help bring the necessary expertise into their immediate family network and help protect the older adults and children of the family from privacy and security risks. However, participants expressed that co-monitoring by extended family members might cause tensions in their families, creating interpersonal conflicts. To alleviate these concerns, participants suggested more control over the privacy features to facilitate sharing their installed apps with only trusted family members. 

\end{abstract}

\begin{CCSXML}
<ccs2012>
   <concept>
       <concept_id>10002978</concept_id>
       <concept_desc>Security and privacy</concept_desc>
       <concept_significance>500</concept_significance>
       </concept>
   <concept>
       <concept_id>10002978.10003029.10003032</concept_id>
       <concept_desc>Security and privacy~Social aspects of security and privacy</concept_desc>
       <concept_significance>500</concept_significance>
       </concept>
 </ccs2012>
\end{CCSXML}

\ccsdesc[500]{Security and privacy}
\ccsdesc[500]{Security and privacy~Social aspects of security and privacy}
%%
%% The code below is generated by the tool at http://dl.acm.org/ccs.cfm.
%% Please copy and paste the code instead of the example below.
%%
% \begin{CCSXML}

% <ccs2012>
%   <concept>
%       <concept_id>10003120.10003130.10011762</concept_id>
%       <concept_desc>Human-centered computing~Empirical studies in collaborative and social computing</concept_desc>
%       <concept_significance>500</concept_significance>
%       </concept>
%  </ccs2012>
% \end{CCSXML}

% \ccsdesc[500]{Human-centered computing~Empirical studies in collaborative and social computing}

%%
%% Keywords. The author(s) should pick words that accurately describe
%% the work being presented. Separate the keywords with commas.
\keywords{Community Oversight; Collaborative Approach; Joint Oversight;  Family Online Safety; Mobile Privacy; Digital Privacy; Security}
%%
%% This command processes the author and affiliation and title
%% information and builds the first part of the formatted document.
\maketitle

\section{Introduction}
A Pew Research study reported that 85\% of U.S. citizens own smartphones ~\cite{nw_demographics_nodate}, and 77\% of them have downloaded and installed different third-party mobile apps on their devices ~\cite{noauthor_majority_2015}. Mobile apps collect personal information (e.g., contact data, emails, photos, location, calendar events, and even browser history) from users when granted permission to do so \cite{nw_mobile_2015}, creating digital privacy threats when this personal information is misused \cite{reardon_50_2019, calciati_automatically_2020}. Unfortunately, the majority of U.S. adults lack knowledge regarding how to protect their digital privacy and security, which increases the potential for privacy and security violations \cite{vogels_americans_2019}. Due to the lack of mobile privacy knowledge at an individual level, networked privacy researchers (e.g., \cite{chouhan_co-designing_2019, aljallad_designing_2019, das_role_2015}) have suggested adopting more collaborative and community-based approaches for managing digital privacy and security, where trusted community members (e.g., family, friends, co-workers) can work together to help keep one another safe online. Interestingly, some research has even shown how adult family members often rely on younger generations of their family (e.g., their teens) for technology support, as youth may be tech-savvier than their parents \cite{works_you_2019, correa_brokering_2015}. 

For instance, a recent study \cite{akter-from-2022} examined family online safety and privacy management by asking parents and teens to evaluate a collaborative mobile app to understand whether parents and teens can help one another manage their mobile privacy and online safety. Despite the hierarchical tensions and asymmetry in privacy and security knowledge between the parents and teens, they found value in such bi-directional joint family oversight for keeping both parties safer online. Meanwhile, our current work acknowledges that a family often consists of other relationships beyond just parents and teens; therefore, it is unclear whether this collaborative approach could be generalized for the whole family by going beyond the parent-teen dyadic relationship. As such, we build upon prior work with parents and teens and extend it by exploring whether, how, and why joint family oversight that includes extended family members may work given a similar use case. Therefore, we first developed a mobile app for Community Oversight of Privacy and Security ("CO-oPS"). Second, we had 19 parents and teens install and evaluate the app in a lab-based setting. In addition to understanding the dynamics of using the app within the context of the parent-teen relationship, participants also expressed the value of including additional family members within the app for broader oversight within an extended family network. As such, we were able to answer the following high-level research questions:

% People to whom the joint family approach could be extended 
\begin{itemize}
\item\textbf{RQ1:} \textit{With whom else could they see including in their trusted networks when sharing and receiving advice on mobile privacy and security?}
\item\textbf{RQ2:} \textit{What would be the potential benefits and drawbacks of including extended family members in a joint family approach?}
\item\textbf{RQ3:} \textit{What are the important design considerations for designing an app for joint family oversight that includes extended family members?}
\end{itemize}

 %in such a way that the parents-teens could co-manage mobile privacy and security with the other family members, including their extended family.

%To answer these research questions, we conducted a lab-based user study with 19 parent-teen (ages 13-17) pairs, where we had them install the CO-oPs app and use its different features that allowed them not only to review one another's mobile privacy and security decisions but also to review other family members' apps and permissions. We also asked them to reflect on whether the CO-oPS app would (or would not) be helpful to use with their other family members. 

Overall, parents and teens felt that they would use the CO-oPS app with their close relatives, particularly grandparents, siblings, and cousins (RQ1). Parents and teens both saw benefits in co-managing their mobile privacy and online safety with their extended families as they thought it would help benefit more vulnerable family members (the older adults and children), and get more expert advice from others within the family who are tech-savvier. However, having extended families included in the CO-oPS family network might also cause unwanted tension and interpersonal issues for the parents and teens, e.g., blaming parents for teens' mobile online behavior, disallowing teens' autonomy, unwanted questioning, and family arguments (RQ2). To alleviate these concerns, they suggested implementing more controls within the privacy feature so that they can share their installed apps with only specified family members (RQ3). Our research makes important contributions to the networked privacy research community by examining whether co-managing mobile privacy, online safety, and security with other family members, beyond just the parent-teen relationship, could potentially help keep entire families safe online. 

% Here, we made the following unique research contributions: 1) Through presenting fictional extended family members on the CO-oPS app, we inspired parents and teens to imagine co-managing mobile online safety and privacy with their other family members; 2) We gained empirical insights into the potential benefits of co-managing their mobile privacy and security with family members, and also the concerns that might make this family joint oversight with extended family challenging to implement; and 3) We provided design-based recommendations toward privacy features that would better support parents and teens in exchanging joint oversight with their extended family members. 

\section{Background}
We place our study within two main streams of research: 1) mobile privacy and security management on the individual level, and 2) collaborative approaches for mobile privacy, safety, and security.
% This section synthesizes prior literature on individual and collaborative management of mobile privacy and security, and also show how our work moves from joint family oversight from just parent-teen to the other relationships of the families.

\subsection{Mobile Privacy and Security Management at the Individual Level}
The proliferation of the usage of smartphones and mobile applications \cite{nw_demographics_nodate} caused mobile phone users to be over-exposed to digital privacy, safety, and security threats ~\cite{vogels_americans_2019, nw_mobile_2015}, as mobile apps get access to users' sensitive and personal information via different permissions ~\cite{vogels_americans_2019, nw_mobile_2015}. Ironically, mobile app users often do not fully understand what these mobile app permissions do and what they are used for ~\cite{ferreira_securacy_2015, alsoubai_permission_2022, park_privacy_2022, ghaiumy_disclose_2021}. Users also lack the understanding of how their personal data are being used by these third parties ~\cite{ghaiumy_disclose_2021}. What is worse, third-party mobile apps may even get unauthorized access to users' information ~\cite{calciati_automatically_2020, reardon_50_2019}. For example, Calciati et al. \cite{calciati_automatically_2020} found that when users give single permission, the app can silently obtain further permissions. They further revealed that many third-party apps leaked and misused users' sensitive data, such as the user's precise location, list of contacts, history of phone calls, and emails, the permissions which users never explicitly granted. Despite these various privacy threats, the majority of US adults lack significant knowledge regarding digital privacy and security and therefore, find it difficult to manage their own privacy and security ~\cite{vogels_americans_2019}. In the next section, we synthesize the relevant literature that suggests adopting collaborative approaches to help resolve individual's challenges in privacy and security management.

\subsection{Collaborative Approaches for Mobile Privacy, Safety, and Security}
Several networked privacy research studies have demonstrated that individuals often seek help and informal advice from their trusted community (e.g., families, friends, coworkers) regarding digital privacy and security ~\cite{rader_stories_2012, rader_identifying_2015}. Users are also influenced by others' privacy and security practices to make changes to their own privacy behaviors \cite{schechter_learning_2015,felt_android_2012, kelley_privacy_2013, das_role_2015, das_effect_2014}. Therefore, networked privacy researchers have called for more collaborative approaches within an individual's networks so that people can exchange mobile privacy, safety, and security support with one another \cite{aljallad_designing_2019, chouhan_co-designing_2019, das_role_2015}. 

However, Kropczynski et al. \cite{kropczynski_towards_2021} have reported that in order to get effective support for digital privacy and security management, one must have some technical expertise in their community. Meanwhile, studies have indicated that teens often provide informal tech support to their family members ~\cite{correa_brokering_2015, kiesler_troubles_2000}. In a recent study, Akter et al. \cite{akter-from-2022} examined a collaborative family oversight approach that allowed teens and their parents to help one another manage their mobile online safety and privacy. Their lab-based study revealed that parents and teens overall valued such collaborative approaches to manage their online safety and mobile privacy. However, there were some tensions between parents and teens because of the differences in hierarchical power and tech-savviness. This work gave us the idea to further investigate whether such joint family oversight mechanisms would be applicable to the other relationships in the family. Building upon these prior studies, we explore whether an app created for Community Oversight of Privacy and Security ("CO-oPS") \cite{chouhan_co-designing_2019} could be beneficial for immediate and extended families to work together to help one another manage their mobile online safety, privacy, and security.

\section{Methods}

\begin{figure*}
%\centering -> This is irrelevant because of the '.5\textwidth' as Mico advised below.

\begin{subfigure}[t]{.234\textwidth}\centering
  \includegraphics[width=\columnwidth]{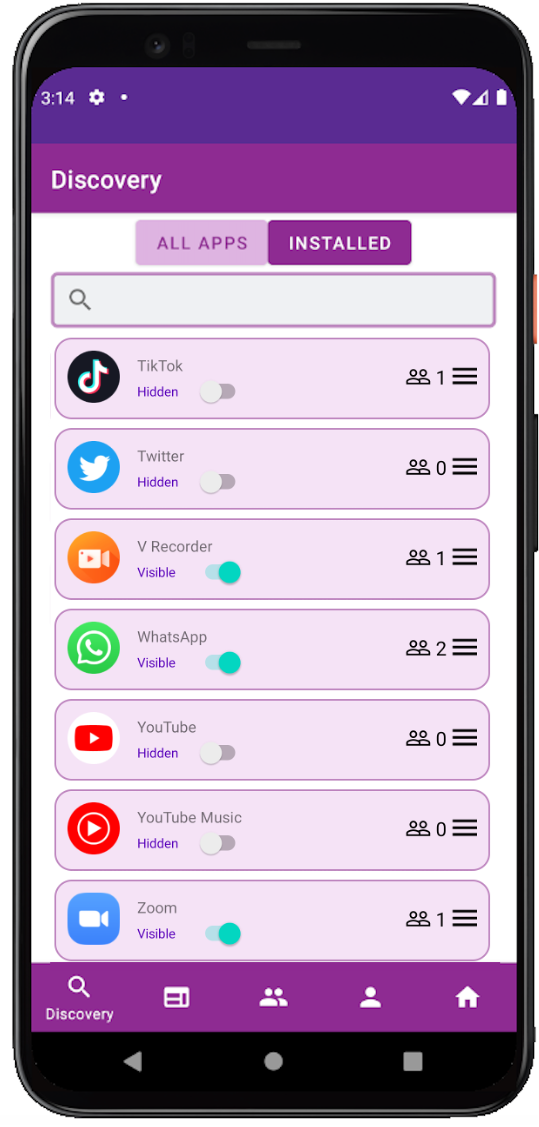}
  \caption{}
\end{subfigure}%
\begin{subfigure}[t]{.239\textwidth}\centering
\centering
 \includegraphics[width=\columnwidth]{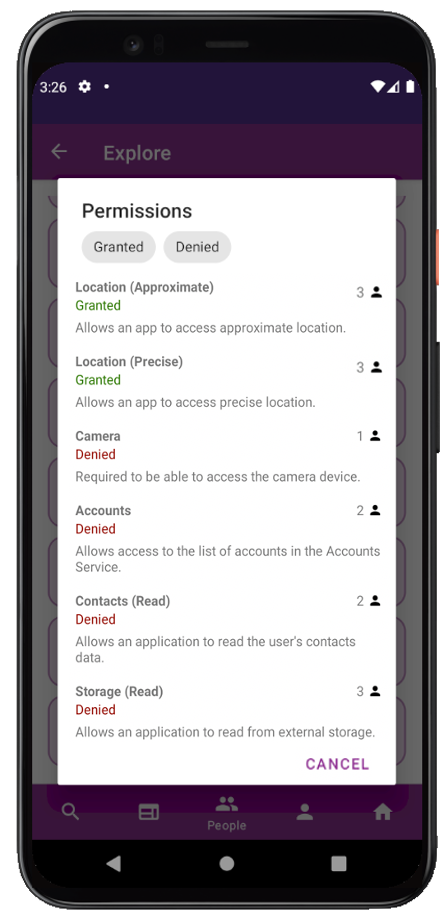}
  \caption{}
\end{subfigure}
\begin{subfigure}[t]{.235\textwidth}\centering
\centering
 \includegraphics[width=\columnwidth]{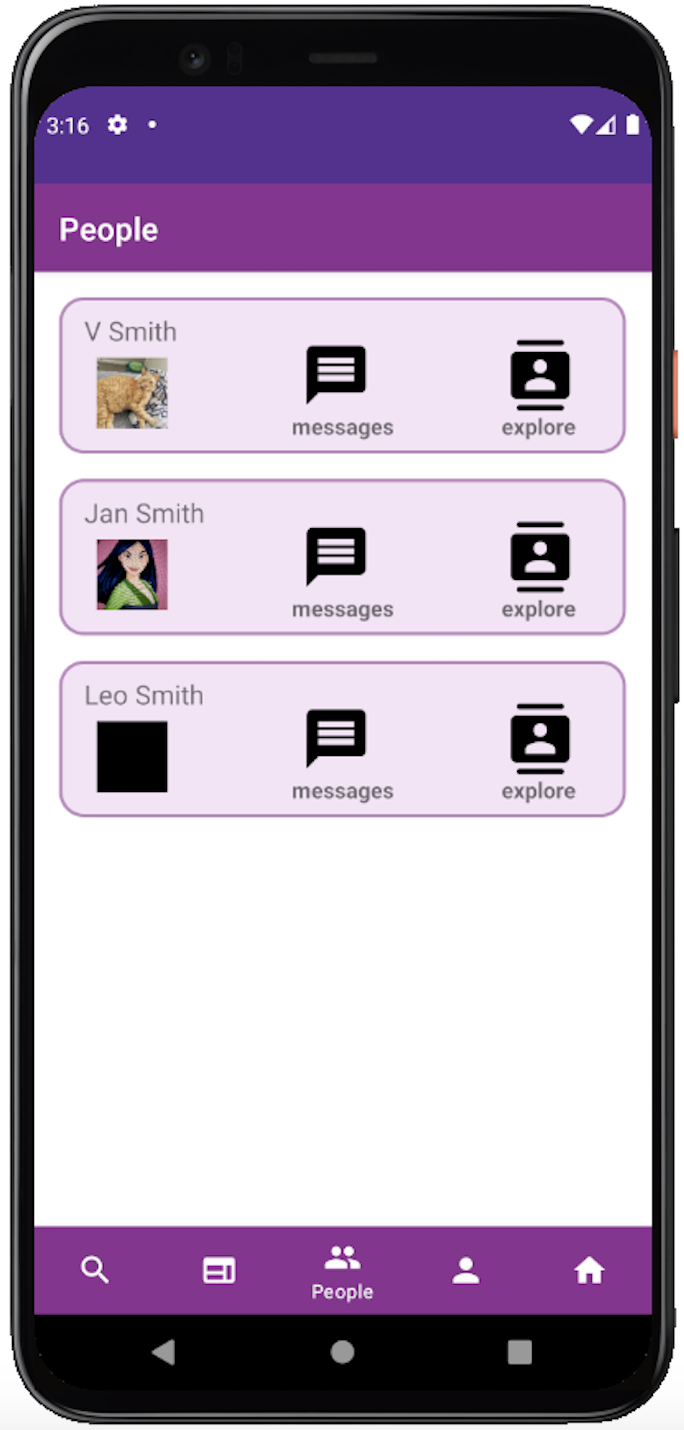}
  \caption{}
\end{subfigure}
\caption{CO-oPS features: (a) Discovery of Installed Apps, (b) Permissions Granted/Denied, (c) People in the Family }~\label{fig:figure1}
\Description[CO-oPS app features]{Figure 1(a) shows the list of apps installed with the hide/show feature, Figure 1(b) shows the permissions granted/denied to an app, and Figure 1(c) shows the list of family members with private message feature.}
% \vspace*{-5pt}
\end{figure*}

\subsection{Design of the CO-oPS App}
We developed the CO-oPS app \cite{akter_CO-oPS_2022} based on the model of community oversight for privacy and security initially proposed by Chouhan et al. in ~\cite{chouhan_co-designing_2019}. This model suggests mechanisms for trusted communities to review one another's mobile privacy and security practices (apps installed and permissions granted) and exchange guidance. The CO-oPS app includes three key aspects: 1) discovery of installed apps, 2) permissions granted/denied, and 3) people in the family. The Discovery feature (Figure-\ref{fig:figure1}a) allows users to review the list of installed apps on their own phone with the ability to hide some apps from their family members. The permissions feature (Figure-\ref{fig:figure1}b) allows users to review the permissions granted or denied to each of the apps installed. In the People screen (Figure-\ref{fig:figure1}c), users can view the list of their family members with the ability to directly message them and explore their installed apps. Here, to help the parents and teens imagine using this app with their other family members, we added two fictional family members: the teen's uncle and aunt. Therefore, during the study, each participant viewed three people in this family list.

\subsection{CO-oPS Parent-Teen User Study}
Our study consisted of two distinct phases: 1) A guided think-aloud exploration of the CO-oPS app with probing questions, and 2) A semi-structured interview with parents and teens, with one component where we asked them to reflect on whether using this with the other family members (including extended family) would be useful for them. The study session started with showing participants a video demo of the CO-oPS app to explain its core functionalities. The participants were then asked to install the app on their phones and use its different features, as shown in Figure-\ref{fig:figure1}. We asked probing questions to learn about participants' reactions to the fictional family members presented on the CO-oPS People page, as well as their suggestions as to who else they would prefer to have in their CO-oPS family network. The presentation of these two extended family members motivated our participants to think about the potential benefits of allowing their extended family to participate in their joint family oversight and encouraged them to think about their concerns of having these family members included in reviewing their apps' permissions. Thus, participants could consider both benefits and drawbacks, which also helped them to brainstorm design suggestions that can support the benefits and alleviate the drawbacks of having extended family included in the CO-oPS family network.  

The study sessions took place on Zoom and were audio and video recorded. We then transcribed the recordings and conducted a grounded thematic analysis for insights, using Braun \& Clarke's \cite{braun2006using} six-phase framework. Overall, we recruited a diverse sample of 19 parent and teen pairs where 42\% of the participants were Asian, 32\% Caucasian, 21\% Hispanic / Latino, and 5\% African American families. 53\% of the teens self-reported as females, and 47\% were males, whereas 58\% of the parents were female, and 43\% were male. The teens' ages ranged from 13-17, with a mean age and standard deviation of 15.4 and 1.4, respectively. The parents’ mean age and the standard deviation were 47.7 and 4.76, respectively, where their ages ranged between 40 to 55.

\section{Results}
In this section, we present the themes that emerged from our qualitative analysis regarding the inclusion of extended family in CO-oPS. Participants' quotations are identified by their IDs (for teens: T1, T2,..T19 and parents: P1, P2,..P19), age, and gender information. 
 
\subsection{People to whom the joint family approach could be extended (RQ1)}
\textit{Most of the parents and teens felt that they would extend the joint family oversight approach to their grandparents, siblings, and cousins.}
%While parents expressed their interest to co-manage mobile privacy mostly with their parents only, . 
More than two-thirds of the parents (68\%, N=13) and half of the teens (53\%, N=10) said they would be interested to have their \textbf{parents (teens' grandparents)} in their CO-oPS family network since they are the closest ones in their family who provide them support, care, and acceptance. 

\begin{quote}
    
% \end{quote}
\textit{"This would be very useful if you include your grandparents. They'd probably freak out if they knew all the apps that are sharing their location...they can then rely on us to help them with the permissions... It's not just they are family, they care about us, they love our kids."} – P9, Female, 51 years old
\end{quote}

Teens showed more flexibility in terms of including other family members, such as siblings, and cousins. About half of the teens (58\%, N=11) and a few parents (16\%, N=3) said they would like to include their \textbf{siblings and cousins (parents' other children, nephews and nieces)} into their CO-oPS family network. We noticed that teens showed more enthusiasm in co-managing with their siblings and cousins because they are of similar age and they get along with them easily. A good number of parents (42\%, N=8) also said they would like to include their own \textbf{siblings (teen's uncles and aunts)} as they are close to their immediate family. Interestingly, around one-third of the parents (32\%, N=6) mentioned they would monitor their \textbf{significant other's} mobile apps and permissions as they are not much aware of the importance of using safe apps and granting safe permissions.

\subsection{Potential benefits and drawbacks of including extended family (RQ2)}
Parents and teens, in general, envisioned benefits in the joint family oversight with their other family members but also saw some potential concerns for including extended family who are not as close to them.

\textit{Most parents and teens thought such joint family oversight would be beneficial for the vulnerable people of their families.} More than half of the participants (63\%, N=12 parents and 47\%, N=9 teens) mentioned that the CO-oPS app would \textbf{help the older adults}, as they often are less aware of the digital privacy threats and also lack the knowledge to monitor or manage app permissions. Both the parents and teens thought that through the CO-oPS app, they would be able to help the older adult family members by letting them know about their unsafe apps and permissions. Next, 58\%, N=11 teens felt it would \textbf{benefit younger children} as they also tend to be less aware of mobile privacy and security issues, similar to the older adults. 

\begin{quote}
    
% \end{quote}
\textit{"A lot of older people like my parents, or like the old people in general, they do not get much time, they are not very tech-savvy either... If my parents, they're included in our app like that,...it would help them.”} – P8, Female, 46 years old

% \begin{quote}
    
\end{quote}
% \subsection{Potential benefits and drawbacks of co-monitoring with extended family (RQ2)}

% all parents and teens discussed some potential benefits of co-monitoring with extended family, also raised concerns about it.

% \subsubsection{Help vulnerable people in the family} more than half of the parents (63\%, N=10) and (47\%, N=9) of teens believed that co-monitoring online safety and privacy with extended family help older adults who are busy to mange their app permissions by themselves, or who are less technologically savvy. As the extended family members can provide additional support and guidance for older adults or less technologically savvy family members, helping them to better understand the risks and potential consequences of their online activity.
% \begin{quote}
% \edit{\textit{"I really see a lot of value, for example, a lot of older people like my parents, grandparents, they do not get much time. They are not very tech-savvy either. So, if they're included in an app like that, that would be really helpful for them.”} – P8}
% \end{quote}

\textit{Parents and teens also saw value in including extended families in the joint family oversight as it would bring necessary expertise into their family network.} Almost half of the teens (47\%, N=9) and a couple parents (11\%, N=2) mentioned that they would be able to \textbf{get more expert advice or guidance from the tech-savvy people} of the extended family. Here, teens mostly mentioned their tech-savvy older cousins and siblings. Similarly, N=2 parents also mentioned their tech-savvy siblings who have better knowledge regarding mobile privacy and security. Apart from this, around one-fourth of the participants (21\%, N=4 parents and 26\%, N=5 teens) said that having their extended family would help them more in terms of being safe online and securing their personal information as more people would be able to \textbf{warn them about unsafe apps}. To this end, they often referred to some popular but controversial social media and gaming apps, e.g., TikTok, Snapchat, Discord, and Instagram, that they came to know from their extended family members. Hence, having more family members in the CO-oPS family network would help them be more aware of different mobile apps and their privacy issues.

\begin{quote}
    
% \end{quote}
\textit{"This would be more about educating the mind and creating awareness because they're [cousins] gonna reevaluate your apps. So we might as well all learn from each other.”} - T14, Female, 16 years old
% \begin{quote}
    
\end{quote}

% \subsubsection{Additional resources and expertise in the family network} Two parents (11\%) and about half of the teens (47\%, N=9) mentioned that co-monitoring with extended family can help in learning from tech-savvy family as they may have more resources or expertise to devote to co-monitoring online safety and privacy, such as additional time, technical knowledge, or experience dealing with similar issues. For example, grandparents, uncles, aunts, and cousins who are more tech savvy may be able to offer advice and guidance on how to use apps safely, or help to monitor and control app usage.
% \begin{quote}
% \edit{\textit{"This would be more about educating the mind and creating awareness, because they're [cousins] gonna reevaluate your apps. So we might as well all learn from each other. I suppose that if I didn't know anything, then I could just ask. They are all tech savvy, so this would help.”} - T14}
% \end{quote}

% \noindent
% some participants (21\%, N=4 parents and 26\%, N=5 teens) also mentioned that involving extended family in co-monitoring can create a sense of greater accountability for all family members, warning each other about unsafe apps and ensuring that all are using apps responsibly.

% \begin{quote}
% \edit{\textit{"I think it could be useful like if there's an app you have installed and maybe your family has also installed that and, someone from your extended family know that the app is just getting users data and selling it, he could like, warn everyone quickly.”}  -P19}
% \end{quote}
\textit{Parents saw additional benefits in including extended family as they would get more help in monitoring their children's mobile online safety.} Around half of the parents (47\%, N=9) felt that involving extended family, especially their siblings (teens' uncles and aunts), in the CO-oPS family network would enable them to \textbf{share the parental responsibilities} to monitor teens' app usage. These parents thought that their siblings would care about their teens as much as they did, and so, they would have some peace of mind knowing there would be other people to monitor their children's mobile online safety. Additionally, one-third of the parents (32\%, N=6) said their children would consider listening to the advice more when they would receive \textbf{guidance and feedback from more people}. A few parents (26\%, N=5) also thought using CO-oPS with extended family would \textbf{strengthen the relationship} between their children and the other family members who live out of their town or state as they would get an opportunity to work together on managing their family online safety and privacy. 
\begin{quote}
    
% \end{quote}
\textit{"Our extended family is pretty concerned about this stuff. So, this is a lot easier to kind of check on our kids, you know to make sure they are not using anything dangerous, and I would know that there are others to tell them about if any safety concerns they might have.”} - P2, Female, 50 years old
% \begin{quote}
    
\end{quote}

\textit{Participants felt that including their extended family in the joint family oversight would bring more tensions.} Almost half of the parents (47\%, N=9) and one-fourth of the teens (26\%, N=5) expressed that involving the extended family in co-managing their mobile privacy and security might become \textbf{more stressful for parents}. They felt that the extended family might blame parents for teens' unsafe online safety behaviors, especially if they find inappropriate mobile apps on teens' phones. Interestingly, 43\%, N=8 teens said such co-managing mobile privacy and security would cause more harm for them because parents might get provoked to \textbf{pull away their autonomy} in using social media, the internet, or even mobile phones. They often brought up the restrictive and authoritarian parenting styles in their uncles' and aunts' families, where their cousins are not allowed to own mobile phones, and this might influence their parents to adopt such restrictions in their family as well. A few participants (16\%, N=3 parents and 5\%, N=1 teens) specifically said that they \textbf{did not see any necessity} of including their extended family as they are not close. 
\begin{quote}
    
% \end{quote}
\textit{"I think once it reaches like extended families, like grandparents and aunts and uncles, it could get a little bit harmful because these are people you aren't living with or aren't as close. I think mostly there would be just drama with my parents, oh, they're doing this or that, you know, stressing them more.”} - T13, Female, 13 years old
% \begin{quote}
    
\end{quote}

\textit{Participants also expressed concerns for potential interpersonal conflicts that might arise.}  More than one-third of the participants (N=42\%, N=8 parents and N=32\%, N=6 teens) envisioned that some people in the extended family might become officious and therefore, there will be some \textbf{unnecessary questioning about their personal choice of app usage}. Some participants (N=37\%, N=7 parents and N=21\%, N=4 teens) also believed that there would be more incidence of \textbf{family arguments when either parents or teens ignore the advice given} regarding their mobile apps installed or permissions granted. 
\begin{quote}
    
% \end{quote}
\textit{“But if it's for the apps, and then you’ll hear like why do you have this app for, then I don't think you have the right to do it [questioning] because they're not completely as close as my immediate family. It's like giving them a new scope to roast me for using a particular app.”} -T16, Female, 14 years old 
% \begin{quote}
    
\end{quote}

\subsection{Design suggestions for a joint family oversight app (RQ3)}
\textit{Participants suggested more control over the app privacy features to allow only specific people to co-monitor their apps and permissions.} Around half of the parents (47\%, N=7) and one-fourth of the teens (26\%, N=5) mentioned that they would want to keep their apps visible only to their immediate families and \textbf{to some specific people} of their extended families, due to the drawbacks highlight above. A couple of teens (11\%, N=2) also said they would want the ability to \textbf{hide from their dominating older siblings}. Interestingly, about one-third of the parents (32\%, N=6) mentioned that they would like to \textbf{hide their teens' apps} from their extended family to avoid any potential tensions or conflicts from happening. Parents often said that they would either manually take their teens' phones or ask their teens to hide their installed apps from the extended family.
\begin{quote}
    
% \end{quote}
\textit{“If this went beyond my immediate family, my husband, and kids, then I would probably reconsider that thought. I would not want all of them to check my things, I would hide apps from them... But again, if there is any option to keep it shown for just a few people from the extended family, not all of them, that would be nice.”} – P12, Female, 55 years old
% \begin{quote}
    
\end{quote}
Additionally, a few parents (16\%, N=3) and teens (21\%, N=4) suggested additional features that would allow them to \textbf{remotely change the apps and permissions} on their extended family members' phones. These participants mostly wanted such features to help some of their relatives who are non-tech savvy and do not live with them. To further explain, they often said that being able to review and let them know would not be enough actually to help them. As some of their extended family, e.g., grandparents, do not have any technical knowledge, they would not be able to follow others' feedback and go to the settings to deny permissions. Therefore, these parents and teens expressed that they would like to have the ability to remotely change the apps and permissions on their family members' phones. 
\begin{quote}
    
% \end{quote}
\textit{“If we could use that with like my grandparents, I would want to change from here [CO-oPS], because they may not know how to go into the place and they can't change it. If you're able to do it here. That would be a thing like I needed in our situation.”} - T3, Female, 14 years old
% \begin{quote}
    
\end{quote}

\section{Discussion}
% In this section, we discuss the major implications of our findings and resulting design decisions.

One of the key lessons learned was that our participants did not implicitly envision their whole family as their CO-oPS family network. They instead envisioned using this family oversight app with people in the family who have strong bonds with them. This reconfirms one of the findings of Chouhan et al.'s paper \cite{chouhan_co-designing_2019}, where their participants were primarily motivated to help only close people. Also, while opening the CO-oPS family network to others in the extended family, our participants were also keenly aware of the trade-off between getting more help versus being able to avoid potential arguments and interpersonal conflicts. As a result, participants tried to negotiate this tension by filtering out the visibility of their apps based on whether someone was close to them or not. Therefore, a key takeaway of our work is to make sure the family members have full agency over with whom their apps are being shared. This desire for more controls is in contrast with the results from Akter et al. \cite{akter-from-2022}, where participants saw little use of privacy features within just the parent-teen context \cite{akter-from-2022}.

On the other hand, our participants wanted to include their extended family in the joint family oversight not only to receive expert advice from others, but also to provide help as well \cite{badillo2020towards}. For instance, parents and teens were willing to include tech-savvy family members in their CO-oPS family network who can help them with privacy and security advice and guidance. This was because they felt the needed expertise that did not exist within their immediate family. In Akter et al.'s work \cite{akter-from-2022}, they found that teens did not trust the feedback of their parents, as they perceived their parents as less tech-savvy. Therefore, including the extended family might provide them with the dependable expert advice and guidance that teens need. Our participants also showed equal enthusiasm to provide help to other family members, especially to older adults who are less tech-savvy. In Kropczynski et al.'s studies \cite{kropczynski_examining_2021, kropczynski_towards_2021}, they found that older adults tend to be on the receiving end of tech support \cite{works_you_2019}, whereas younger adults are more likely to be tech support givers. So, joint family oversight could support such tech caregiving mechanisms, including the support for mobile privacy and security. Akter et al. \cite{akter-from-2022} reported in their study that teens did not feel empowered to monitor their parents' apps and permissions and hence, were reluctant to participate in joint family oversight. So, teens' being interested to provide help to others in the extended family, such as their grandparents, shows the potential for increasing their engagement in the joint family oversight mechanism. 

\section{Limitations and Future Research}
We recognize several limitations of our study. First, we examined the perceived benefits and drawbacks of using the CO-oPS app with extended family members only from the parents' and teens' perspectives, as this was an emergent finding from our primary study that was worthy of further investigation. As such, it would be an important next step to explore extended family members' opinions in future work. Another potential limitation was that we presented the teens' uncle and aunt as their fictional extended family members on CO-oPS, which might have led participants to think more about the prospective problems that might arise from co-managing with these family members. Interestingly, however, our participants felt that they would use the CO-oPS app mostly with their other relatives: teens' grandparents, siblings, and cousins. Lastly, because the nature of our study was lab-based, parents and teens could not evaluate the CO-oPS app in a realistic setting; thus, in future studies, we would want to deploy the CO-oPS app among groups of family members, including diverse family relationships. 

\section{Conclusion}
Given the continued proliferation and usage of smartphones and third-party mobile apps, we believe community members can work side by side to co-manage their mobile privacy, safety, and security. Our study explored a joint family approach, highlighting parents' and teens' perceived benefits, concerns, and design considerations for co-managing their mobile privacy and security with other members of their extended families. Our work demonstrates the added benefits and challenges when broadening an oversight community beyond parents and teens. Generally, participants felt that such a collaborative approach with extended family would help them exchange privacy and security support with those they cared about. Yet, potential tensions and interpersonal conflicts would require additional controls on collaborative monitoring so that only trusted and close family members can review their mobile privacy and security behaviors. We will continue to build upon this work to examine how we can help people successfully co-manage mobile privacy, safety, and security within their families.

\begin{acks}
We acknowledge the contributions of Nazmus Sakib Miazi, Nikko Osaka, Anoosh Hari, and Ricardo Mangandi in developing the CO-oPS application. We would also like to thank the parents and teens who participated in our study. This research was supported by the U.S. National Science Foundation under grants CNS-1844881, CNS-1814068, CNS-1814110, and CNS-1814439. Any opinions, findings, conclusions, or recommendations expressed in this material are those of the authors and do not necessarily reflect the views of the U.S. National Science Foundation.
\end{acks}

%%
%% The next two lines define the bibliography style to be used, and
%% the bibliography file.
\bibliographystyle{ACM-Reference-Format}
\bibliography{sample-base}

\newpage

% \appendix

% \label{Appendices}

% \begin{appendices}
% \section{}

% % \section{Some More Notation}
% \end{appendices}

% \end{appendices}
\end{document}